\documentclass[aps, 10pt, preprintnumbers, prd, amsmath, amssymb,twocolumn, notitlepage, nofootinbib]{revtex4} 

\usepackage{graphicx}
\usepackage{bm}
\usepackage{subfigure}
\usepackage{epsfig}
\usepackage{color}
\usepackage{mathtools}
\usepackage{cases}
\usepackage{booktabs}
\usepackage{amsmath}

\usepackage[
colorlinks=true,
filecolor=black,
anchorcolor=blue,
linkcolor=cyan,
citecolor=cyan, 
urlcolor=cyan,
linktocpage=true,
plainpages=false,
breaklinks=true,
pdfstartview=FitH
]{hyperref}

\newcommand{\sbh}{\sigma_{\r{bh}}}
\newcommand{\mbh}{m}
\newcommand{\neff}{N_{\rm{eff}}}
\newcommand{\dneff}{\Delta N_{\rm{eff}}}
\newcommand{\ms}{m_{\odot}}
\newcommand{\fbh}{f_{\r{bh}}}
\newcommand{\GW}{\Omega_\r{GW}}
\newcommand{\rd}{\r{d}}
\newcommand{\ps}{P_{\mathcal{R}}}

\newcommand{\be}{\begin{equation}}
\newcommand{\ee}{\end{equation}}
\newcommand{\bea}{\begin{equation}\begin{aligned}}
\newcommand{\eea}{\end{aligned}\end{equation}}
\DeclareRobustCommand{\r}[1]{{\rm #1}}
\DeclareRobustCommand{\Eq}[1]{Eq.~(\ref{#1})}
\DeclareRobustCommand{\Fig}[1]{Fig.~\ref{#1}}

\begin{document}

\title{Implications of Pulsar Timing Array Results for High Frequency Gravitational Waves}

\author{Junsong Cang$^{1}$}
\email{cangjunsong@outlook.com}
\author{Yu Gao$^{2}$}
\email{gaoyu@ihep.ac.cn}
\author{Yiming Liu$^{3}$}
\email{7520220161@bit.edu.cn}
\author{Sichun Sun$^{3}$}
\email{sichunssun@bit.edu.cn}

\affiliation{$^1$ School of Physics, Henan Normal University, Xinxiang, China}
\affiliation{$^{2}$Key Laboratory of Particle Astrophysics, Institute of High Energy Physics,
Chinese Academy of Sciences, Beijing 100049, China}
\affiliation{$^{3}$School of Physics, Beijing Institute of Technology, Beijing, 100081, China}

\begin{abstract}

Several pulsar timing array (PTA) experiments such as NANOGrav and PPTA recently reported evidence of a gravitational wave (GW) background at nano-Hz frequency band. This signal can originate from scalar-induced gravitational waves (SIGW) generated by the enhanced curvature perturbation. Production of SIGW is expected to be accompanied by formation of primordial black holes (PBH), which can emit GW through binary mergers. Here we perform a joint likelihood inference on PTA datasets in combination with existing limits on PBH abundance and GW density, we derive full Bayesian posteriors for PBH distribution and relevant PBH merger signal. Our results show that analysis using PTA data alone implies significant overproduction of PBHs, and accounting for current PBH limits causes visible shifts in SIGW posterior. If PTA signals are indeed of SIGW origin, the required curvature perturbation amplitude produces PBHs in a narrow mass window of $[6 \times 10^{-2}, 2 \times 10^{-1}]\ m_\odot$. Mergers of these PBHs can leave a strong GW signature in $[10^{-3}, 10^5]$ Hz frequency range, to be detectable at upcoming interferometers such as LISA, aLIGO, Einstein Telescope, DECIGO and BBO, etc. This offers a multi-frequency opportunity to further scrutinize the source of the observed PTA signal and can potentially improve current PBH constraints by up to 5 orders of magnitudes.

\end{abstract}

\maketitle

\section{Introduction}
Enthusiasm in primordial black holes (PBHs)~\cite{Zeldovich:1967lct,Carr:1974nx} has grown immensely after the LIGO discovery of gravitational wave signals~\cite{LIGOScientific:2018mvr} in agreement with merger events of black holes above stellar masses~\cite{Bird:2016dcv}.
With sufficiently large primordial curvature perturbation $\mathcal{R}$,
such black holes can be produced in the early Universe from gravitational collapse of overdense regions.
PBHs are under extensive searches for their astrophysical signals
~\cite{Cang:2021owu,Wang:2019kaf,Khlopov:2008qy,Belotsky:2014kca,Belotsky:2018wph,Choudhury:2013woa,Choudhury:2023vuj,
Choudhury:2023jlt,Choudhury:2023rks,Choudhury:2023hvf,Wang:2020uvi},
see Ref.~\cite{Carr:2021bzv} for a recent review.
The curvature perturbation $\mathcal{R}$ which may source PBH production is also predicted to produce scalar-induced gravitational waves (SIGW)~\cite{Domenech:2021ztg,Cang:2022jyc},
which offers a glimpse at valuable information of early fluctuations at late-time gravitational wave detectors.

Recently several Pulsar Timing Arrays (PTAs) observatories including NANOGrav(North American Nanohertz Observatory for Gravitational Waves)~\cite{NANOGrav:2023gor}, 
Chinese PTA (CPTA)~\cite{Xu:2023wog},
Parkes PTA (PPTA)~\cite{Reardon:2023gzh}
and European PTA (EPTA)~\cite{EPTA:2023gyr,EPTA:2023fyk}
reported strong evidence for gravitational wave (GW) background at nano-Hertz waveband,
which verifes previous claims~\cite{NANOGrav:2020bcs, Chen:2021rqp, Goncharov:2021oub, Antoniadis:2022pcn}.
These observations incited extensive studies on potential sources of the observed 
GW background~\cite{Franciolini:2023pbf, 
Franciolini:2023wjm, Liu:2023ymk, Ellis:2023dgf, Wu:2023hsa, Li:2023yaj,Sun:2021yra,Li:2023vuu, Battista:2021rlh,
DeFalco:2023djo,Konoplya:2023fmh,Ben-Dayan:2023lwd,Balaji:2023ehk,Kohri:2020qqd,Inomata:2023zup,Vagnozzi:2020gtf,Benetti:2021uea,
Vagnozzi:2023lwo,Guo:2023hyp,Oikonomou:2023bli,Oikonomou:2023qfz,Choudhury:2023kam,Choudhury:2023hfm,Bhattacharya:2023ysp,Madge:2023cak},
including supermassive black holes~\cite{NANOGrav:2023hvm,Middleton:2020asl,NANOGrav:2023hfp,EPTA:2023xxk}, 
merging PBHs~\cite{Depta:2023qst, Gouttenoire:2023nzr},
phase transitions~\cite{Bian:2020urb,NANOGrav:2021flc,Xue:2021gyq,Wang:2022wwj} and 
axion topological defects~\cite{Wang:2022rjz,Ferreira:2022zzo,Inomata:2023drn}, etc.

As a viable nano-Hz source, an SIGW-emitting overdensity collapse process generally requires a curvature power spectrum $\ps$ with a large amplitude at the scale of $k\sim 10^{8}$ Mpc$^{-1}$~\cite{Franciolini:2023pbf,NANOGrav:2023hvm,Inomata:2023zup} to be consistent with the PTA data.
Although $\ps$ is tightly constrained at large scales ($k \lesssim 10 \r{Mpc^{-1}}$) by observations of cosmic microwave background (CMB) and large scale structures~\cite{Hunt:2015iua, Planck:2018vyg}, 
at smaller scales $\ps$ remains relatively poorly constrained~\cite{Cang:2022jyc,Inomata:2018epa,Choudhury:2023kdb} and can potentially assume the amplitude amplitude required to explain PTA signals.

In this paper,
we show that high-frequency GWs are predicted from an SIGW generating $\ps$ amplitude in consistency with PTA data. 
A perturbation spectrum that yields the required nano-Hertz gravitational wave will also lead to the formation of PBHs with mass in $[6\times 10^{-2}, 2 \times 10^{-1}] m_\odot$ range, 
and the mergers of these PBHs produce a GW background peaked at around MHz frequencies~\cite{Inomata:2023zup}. 
Such high-frequency GW signals shall be readily detectable at various upcoming observatories such as Einstein Telescope (ET)~\cite{Punturo:2010zz},
Deci-Hertz Interferometer Gravitational Wave Observatory (DECIGO)~\cite{Kawamura:2020pcg} and Big Bang Observer (BBO)~\cite{phinney2004big}. 
Since the endeavor of the whole gravitational wave frequency spectrum searches has begun, 
with various proposals already operating in MHz-GHz band\cite{Aggarwal:2020olq, Domcke:2022rgu,Gao:2023gph,Sun:2020gem}, 
cross-linking the ultralow nano-Hz gravitational waves to the ultrahigh-frequency searches in a multi-messenger task in GW spectrum space is intriguing.

This work is organized as follows,
Sec.\ref{dsfhuiyug39dy783eudygfcghjskudtyecs} briefly reviews GW signal from SIGW and merging PBH,
our inference settings are detailed in Sec.\ref{dsfefe3e45rdfghji87ytrds}.
We present our inference results in Sec.\ref{e2f8sanb_sfax34354tghr87t6eftcshidftyyrtdasadwu} and conclude in Sec.\ref{e2f8sanb_sfax34354tghr87t6eftcs_dsfeu}.
We will show that the PBH merger GW in consistency with recent PTA data can be well probed by future interferometry experiments.

\section{GW signal}
\label{dsfhuiyug39dy783eudygfcghjskudtyecs}

As a good representative for a large class of curvature perturbation models,
we consider a log-normal curvature power spectrum~\cite{Inomata:2018epa, NANOGrav:2023hvm, Franciolini:2023pbf, Pi:2020otn, Chen:2021nio} that feature a characteristic perturbation scale,
\be
\ps
=
\frac{A}{\sqrt{2 \pi \Delta^2}}
\exp
\left[
-\frac{(\ln k/k_*)^2}{2\Delta^2}
\right],
\label{dsf9887hdsf}
\ee
where $A,\ k_*$ and $\Delta$ are model parameters, 
which describe the amplitude,
peak location and the width of $\ps$ respectively.

\subsection{Scalar Induced GW}

Upon horizon crossing,
$\ps$ will modify the radiation quadruple moment and generate SIGW at second order,
whose energy density per log frequency interval today is given by~\cite{Cang:2022jyc,Ando:2018qdb,Kohri:2018awv,Inomata:2018epa},
\be
\begin{aligned}
\Omega_\r{GW}
&\equiv
\frac{1}{\rho_{\rm cr}}
\frac{{\rm d}\rho_{\rm GW}}{{\rm d}\ln f}
\\
& = 
0.29~
\Omega_{\r{r}}
\left(
\frac{106.75}{g_{\ast}}
\right)^{1/3}
\\
&
\ \ \ \ 
\times
\int^\infty_0 dv \int^{1+v}_{|1-v|} du \left[ \frac{4v^2-(1-u^2+v^2)^2}{4u^2v^2}\right]^2
\\
&\ \ \ \  \times \left(\frac{u^2+v^2-3}{2 u v}\right)^4 F(u,v) \ps(kv)\ps(ku),
\end{aligned}
\label{dsfees8sjwof84deygshcx}
\ee
\be
\begin{aligned}
F(u,v)
&=
\left( \ln\left| \frac{3-(u+v)^2}{3-(u-v)^2}\right|-\frac{4 u v}{u^2+v^2-3}\right)^2
\\
&
+
\pi ^2 \Theta \left(u+v-\sqrt{3}\right)
,
\end{aligned}
\ee
here $\rho_\r{cr}$ is critical density,
$\rho_\r{GW}$ is energy density in GW,
$\Omega_\r{r}$ is fractional density in radiation,
we assume neutrinos to be massless such that $\Omega_\r{r} =9.1 \times 10^{-5}$.
$g_*$ is total degree of freedom for massless particles when the mode $k$ enters horizon ($k=aH$)~\citep{KolbTurner1990,Wallisch:2018rzj},
$\Theta$ is Heaviside step function,
and frequency $f$ is related to wavenumber $k$ via~\cite{Chen:2021nio},
\be
f = 1.546 \times 10^{-15}\left(\frac{k}{\r{Mpc^{-1}}}\right) \r{Hz}
\label{eq_k2f}
\ee

\subsection{PBH production}

If sufficiently large,
$\ps$ can create overdense regions that gravitationally collapse into PBHs with mass~\cite{Cang:2022jyc,Carr:2009jm,Nakama:2016gzw,Ozsoy:2018flq,Chen:2021nio},
\be
\mbh
=
2.43 \times 10^{-4}
\left(
\frac{\gamma}{0.2}
\right)
\left(
\frac{g_{\ast}}{106.75}
\right)^{-1/6}
\left(
\frac{k}{10^8\rm Mpc^{-1}}
\right)^{-2}
\ms,
\label{k2m_relation}
\ee
here $\gamma$ is collapse efficiency,
for which we adopt a typical value of $\gamma = 0.2$ following
~\cite{
Carr:2009jm,
Ozsoy:2018flq,
Chen:2021nio}.
The corresponding distribution of PBH abundance is given by~\cite{Carr:2020xqk, Young:2014ana,Ozsoy:2018flq,1975ApJ...201....1C},
\be
\begin{aligned}
\frac{\r{d}\fbh}{{\rm d} \ln \mbh}
=
0.28
\left(
\frac{\beta (m)}
{10^{-8}}
\right)
\left(
\frac{\gamma}{0.2}
\right)^{3/2}
\left(
\frac{106.75}{g_{\ast}}
\right)^{1/4}
\left(
\frac{\ms}{m}
\right)^{1/2}
,
\label{9876tyuuyt}
\end{aligned}
\ee
where $\fbh \equiv \rho_\r{bh}/\rho_\r{dm}$ is fraction of DM made of PBHs,
$\rho_\r{bh}$ and $\rho_\r{dm}$ denote mass densities of PBH and DM respectively,
and
\be
\beta (m)
\simeq
\sqrt{
\frac{2 \bar{\sigma}^2}
{\pi \delta_{\rm c}^2}
}
\r{exp}
\left(
-\frac{\delta_{\rm c}^2}{2 \bar{\sigma}^2}
\right),
\label{7865rt54321qw}
\ee
\be
\bar{\sigma}^2
(m)
=
\frac{16}{81}
\int_0^{\infty}
\frac{\r{d} k'}
{k'}
\left(
\frac{k'}{k}
\right)^4
\ps
(k')
\mathcal{T}^2\left(\frac{k'}{k} \right)
W^2\left(\frac{k'}{k} \right),
\label{8765rfgrtf}
\ee
here $\mathcal{T}$ is the transfer function~\cite{Musco:2020jjb,Ando:2018qdb},
\be
\mathcal{T}(x)
=
3
\frac{\sin (x/\sqrt{3}) -(x/\sqrt{3}) \cos (x/\sqrt{3})}
{(x/\sqrt{3})^3}
\ee
and we adopt a real space top-hat form for the window function $W$~\cite{Ando:2018qdb},
\be
W(x)
=
3
\frac{\sin x -x \cos x}
{x^3}.
\label{eq_WindowFunction}
\ee

Finally in \Eq{7865rt54321qw} $\delta_\r{c}$ is the density fluctuation threshold for gravitational collapse.
The value of $\delta_\r{c}$ is dependent on the shape of $\ps$ and choice of window function~\cite{Musco:2020jjb,Dandoy:2023jot,Escriva:2019phb,Young:2020xmk}.
For our $\ps$ and window function,
$\delta_\r{c}$ has been carefully examined in Refs.\cite{Musco:2020jjb,Franciolini:2021nvv,LISACosmologyWorkingGroup:2023njw} through numerical simulation and is shown to range between 0.4 and 0.6.
Throughout this work,
we use the result presented in~\cite{Musco:2020jjb} to compute $\delta_\r{c}$.

\begin{figure}[t]
\centering
\includegraphics[width=8.3cm]{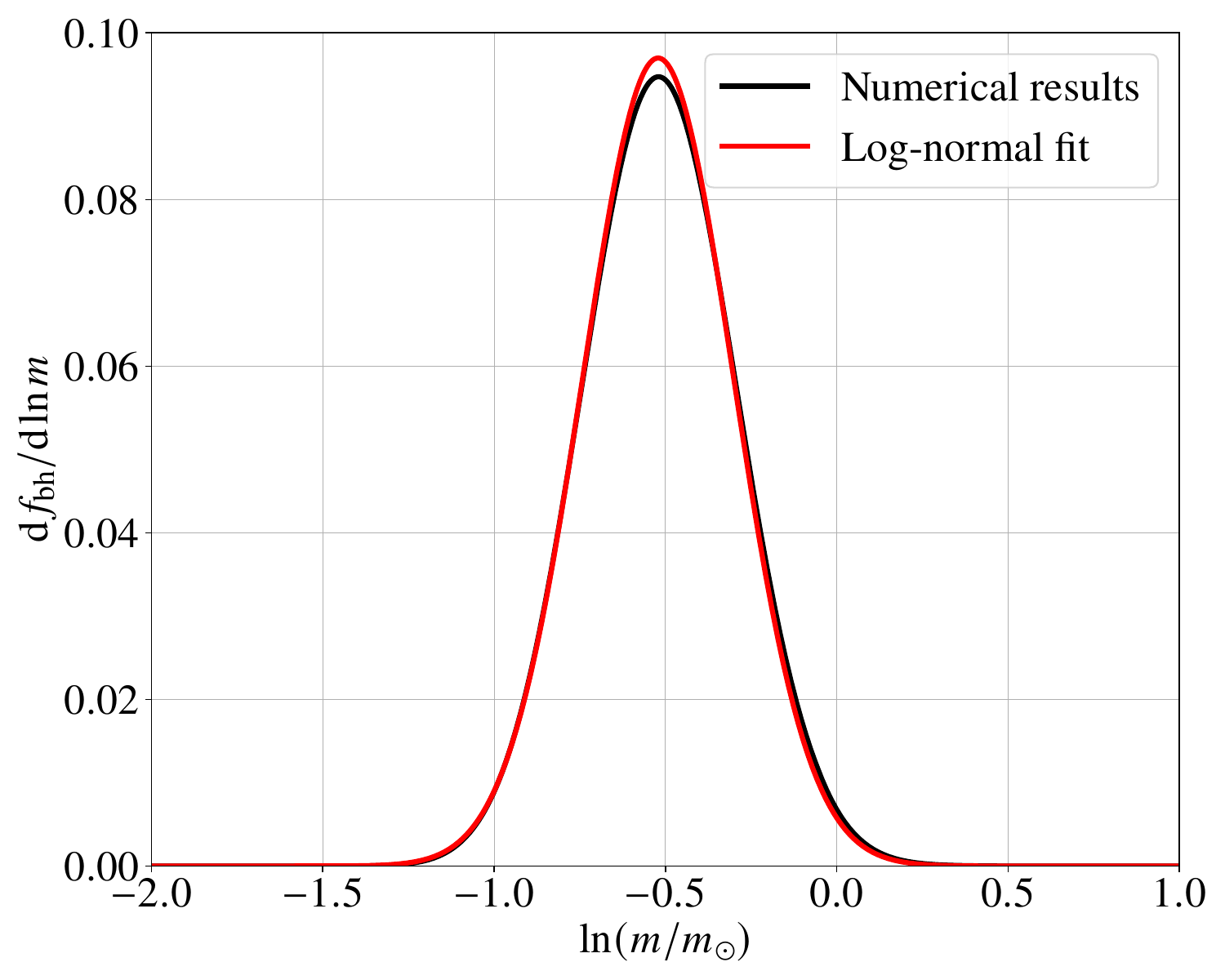}
\caption{
PBH mass distribution function $\rd \fbh/\rd \ln m$ for $\ps$ parameters $A = 7 \times 10^{-4}$,
$\Delta = 0.3$ and $k_* = 10^7\ \r{Mpc^{-1}}$.
The black solid line shows the numerical result computed following Eqs.(\ref{k2m_relation} - \ref{eq_WindowFunction}),
and the red solid line represents the analytic fitting results using the log-normal model in \Eq{dshiuy65}.
}
\label{PBH_MassFunction_Example}
\end{figure}

In practice,
we find that the distribution calculated from \Eq{9876tyuuyt} can be very well fit by a lognormal profile~\cite{Cang:2022jyc},
\be
\psi
\equiv
\frac{1}{\fbh}
\frac{\r{d}\fbh}{\rd \ln \mbh}
=
\frac{1}{\sqrt{2 \pi} \sbh}
\exp
\left[
-
\frac{\r{ln}^2(\mbh/m_{\r{bh}})}
{2 \sbh^2}
\right]
\label{dshiuy65}
\ee
where $m_\r{bh}$ and $\sigma_\r{bh}$ are peak and width of the distribution respectively and we solve their values with a simple least-square fitting method.
In \Fig{PBH_MassFunction_Example} we show an example of distribution functions numerically calculated from \Eq{9876tyuuyt} and that obtained through analytic lognormal fit in \Eq{dshiuy65},
and it can be seen that \Eq{dshiuy65} provides an excellent fit to the actual distribution.
Ref.\cite{Gow:2020cou} proposed alternative fitting models that may provide even better fits,
e.g. a skew-lognormal model can better capture possible skewness in the distribution by introducing an additional shape parameter.
However we find that the lognormal model provides sufficient fitting precision for the purpose of this work,
therefore for the sake of simplicity,
hereafter we will use \Eq{dshiuy65} to parameterise the distribution given by \Eq{9876tyuuyt}.

\subsection{Merger GW}
\label{dsifhuey7t63sd}

After their production, 
PBHs can form binary systems which subsequently merge and emit GW at higher frequencies.
The comoving merger rate $R$ of a pair of PBHs with mass $m_1$ and $m_2$ is given by~\cite{Raidal:2018bbj},
\be
\begin{aligned}
\frac{\rd R}{\rd m_1 \rd m_2}
&\simeq
\frac{1.6 \times 10^6}
{\r{Gpc^3~ yr}}
\fbh^{\frac{53}{37}}
\eta^{-\frac{34}{37}}
\left(
\frac{M}{m_\odot}
\right)^{-\frac{32}{37}}
\\
&
\times
\left(
\frac{t}{t_0}
\right)^{-\frac{34}{37}}
S~ 
\psi(m_1)~
\psi(m_2)
\end{aligned}
\label{dfiiuiu7thxdcyeg63}
\ee
where 
$M = m_1 + m_2$,
$\eta = m_1 m_2 / M^2$ and
$t$ is the time of merger,
$t_0 = 13.8~\r{Gyr}$ is current age of the Universe.
$S$ is a suppression factor and we take its form from Ref~\cite{Raidal:2018bbj, Hutsi:2020sol}:
\be
\begin{aligned}
S
&=
S_\r{late} 
\frac{\r{e}^{-\bar{N}(y)}}
{\Gamma(21/37)}
\int \rd v
v^{-\frac{16}{37}}
\exp
\left[
-
\phi
-
\frac{3 \sigma^2_\r{M} v^2}
{10 \fbh^2}
\right],
\end{aligned}
\label{fdvdsx2rsdcvio}
\ee
\be
S_\r{late} \simeq
\r{min}
\left[
1,
1.96
\times
10^{-3}
x^{-0.65}
\exp
\left(
0.03
\ln^2
x
\right)
\right],
\label{fduf7ngdygr543}
\ee
\be
\phi
=
\bar{N}(y)
\left<
m\right>
\int
\frac{\rd \mbh}{\mbh}
\psi(m)
F
\left(
\frac{M}
{
\left<
m
\right>
}
\frac{v}
{\bar{N}(y)}
\right),
\ee
here $x = (t/t_0)^{0.44} \fbh$,
$\sigma_\r{M} \simeq 0.004$,
$\bar{N}(y)$ is the expected number of PBHs within a comoving radius of $y$ around the binary~\cite{Hall:2020daa}, 
and we take $\bar{N}(y) \simeq
{M \fbh}
/
[
\left<m\right>
(\fbh + \sigma_\r{M})]
$
following~\cite{Raidal:2018bbj,Hutsi:2020sol,Hall:2020daa}. This choice has been shown to be in agreement with numerical simulations for $\fbh \le 0.1$~\cite{Raidal:2018bbj,Hall:2020daa}.
$\left<m\right>$ is the mean of PBH mass over number density distribution~\cite{Hutsi:2020sol},
which equals $m_\r{bh} \r{e}^{-\sbh^2/2}$ for our log-normal mass distribution in~\Eq{dshiuy65}.
$
F(z)
=\ 
_1{\mathcal{F}}_2
(-1/2,3/4,5/4;-9z^2/16)
-1
$,
and $_1{\mathcal{F}}_2$ is the generalized hypergeometric function. 

It can be shown straightforwardly that the energy density parameter $\GW$ (defined in the first line of \Eq{dsfees8sjwof84deygshcx}) for merging PBHs takes the form~\cite{Depta:2023qst},
\be
\GW
=
\frac{f}{\rho_\r{cr}}
\int \frac{\rd z~\rd R}
{(1+z)~H}
\frac{\rd E_\r{GW}(f_\r{r})}
{\rd f_\r{r}}
\ee
here $f_\r{r} = (1+z)f$ denotes the source frequency,
$\rd E_{\r{GW}}(f_\r{r})/\rd f_\r{r}$ is source energy spectrum for each PBH merger event,
for which we adopt~\cite{Zhu:2011bd},
$H = H_0[\Omega_\Lambda + \Omega_\r{m}(1+z)^3 + \Omega_\r{r}(1+z)^4]^{1/2}$ is Hubble parameter,
and we use cosmological parameter values from Planck 2018 results~\cite{Planck:2018vyg}: $H_0 = 67.66\ \r{km s^{-1}Mpc^{-1}}$,
$\Omega_\Lambda = 0.6903$, $\Omega_\r{m} = 0.3096$.

\section{Inference Settings}
\label{dsfefe3e45rdfghji87ytrds}

We analyze the PTA datasets using our SIGW model to map out the credible regions of PBH parameters and associated high-frequency merger signal.
Our likelihood is
\be
\mathcal{L}
\propto
\exp
\left[
-\sum_{i}
\frac{(x_i - u_i)^2}{2\sigma_i^2}
\right],
\label{dshf3765rghfdv}
\ee
here the index $i$ denotes frequency.
We sample GW data in log space,
as such $x$ is $\ln \Omega_\r{GW}$ for our SIGW model,
$u$ is the measured median for $\ln \Omega_\r{GW}$.
The error term $\sigma$ is assumed to be asymmetric,
therefore it takes upper error bar $\sigma_u$ when $x_i > u_i$ and lower error bar $\sigma_l$ when $x_i \le u_i$.
We use the datasets from NANOGrav-15~\cite{NANOGrav:2023gor}, 
IPTA~\cite{Antoniadis:2022pcn} and PPTA~\cite{Reardon:2023gzh} in our inference, and 
we follow Ref.~\cite{Franciolini:2023pbf} and estimate the signal median and error bars for each experiment using the $\Omega_\r{GW}$ posterior summarised in Ref.~\cite{Franciolini:2023wjm,Li:2023yaj}. 
For validation, we have checked that our fitting agrees very well with Ref.~\cite{Franciolini:2023pbf} and~\cite{NANOGrav:2023hvm} when using NAGOGrav-15 data alone,
and we provide a brief comparison of parameter posteriors derived with our approximate GW likelihood with the result from the full NAGOGrav-15 data analysis in appendix.\ref{appendix_Comparison_with_Full_GW_Analysis}.

We combine the datasets from different PTA experiments when computing the likelihood in \Eq{dshf3765rghfdv},
however we note that this procedure is approximate since different PTA observations may have some pulsars in common,
and pulsar noise modeling may not be unified between different experiments.

As a contributor to dark radiation,
the extra energy budget in SIGW will also change the effective relativistic degrees of freedom $\neff$.
In Planck 2018 results (hereafter PLK)~\cite{Planck:2018vyg},
a joint analysis of datasets from CMB, 
baryon acoustic oscillations (BAO) and Big Bang Nucleosynthesis (BBN) presented an upper bound at 95\% confidence limit (C.L.)~\cite{Planck:2018vyg,Cang:2022jyc}
\be
\dneff
\equiv
\neff - 
3.046
\le
0.175
,\ 95\%\ \r{C.L.},
\label{dsfhi38r767fdg}
\ee
here $3.046$ is the value of $\neff$ predicted by the standard model of particle physics~\cite{deSalas:2016ztq,Planck:2018vyg,Cang:2022jyc}. 
This $\neff$ constraint translates into an upper bound on the integrated GW density, 
or $\int \rd \ln f\Omega_{\r{GW}} < 2.1 \times 10^{-6}$~\cite{Cang:2022jyc}.

To accommodate the PLK $\neff$ limits, 
we add a $\dneff < 0.175$ prior to our inference.
Since $\fbh$ in our mass range has been constrained to $\mathcal{O}(0.1)$~\cite{Carr:2020xqk},
we also use a prior of $\fbh \le 0.1$ to ensure that PBHs produced by $\ps$ does not violate the existing abundance constraints.
Given all the discussions above, we consider three benchmark inference settings:
\begin{itemize}
\item {\tt PTA}: Use PTA GW data alone.
\item {\tt PTA+$\dneff$}: Use PTA GW data and PLK $\dneff$ limits in \Eq{dsfhi38r767fdg}.
\item {\tt PTA+$\dneff$+PBH}: Use PTA GW data and PLK $\dneff$ limits along with the prior of $\fbh < 0.1$.
\end{itemize}

\begin{table}[htp]
\begin{tabular}{c|c|c|c}
\hline
Parameters & Prior & 95\% Limits & 95\% Limits\\
 & range & {\tt PTA} & {\tt PTA+$\dneff$+PBH}\\
\hline
$\log_{10}A$& [-5, 3] & [-1.62, 0.71] & [-2.09, -1.85]\\
$\log_{10}(k_*/\r{Mpc^{-1}})$ & [4, 10] & [7.34, 10.05] & [6.91, 7.18]\\
$\Delta$& [0.02, 5] & [0.06, 3.14] & [0.16, 1.03]\\
\hline
$\log_{10}f_\r{bh}$& $\le-1$ & [4.31,  13.19] & [-2.31, -1.00]\\
$\log_{10}[m_\r{bh}/m_\odot]$ & -- & [-10.34, -1.41] & [-1.22, -0.69]\\
$\sbh$ & -- & [0.48, 2.75] & [0.16, 0.50]\\
$\Delta N_\r{eff}$&  $\le 0.175$ & $[0, 7.71] $ & $[1.26, 1.88]\times 10^ {-4}$\\
\hline
\end{tabular}
\caption{
Parameters in our inference and their allowed range and marginalized 95\% C.L. limits.
$A$, $k_*$ and $\Delta$ are our free model parameters,
whereas $\fbh,\ m_\r{bh},\ \sbh$ and $\Delta N_\r{eff}$ are parameters derived from $A$, $k_*$ and $\Delta$.
Note that fitting to PTA data alone significantly overproduces PBHs.}
\label{tabd76543wedfghsa}
\end{table}

\begin{figure*}[htp]
\centering
\subfigbottomskip=-500pt
\subfigure{\includegraphics[width=8.5cm]{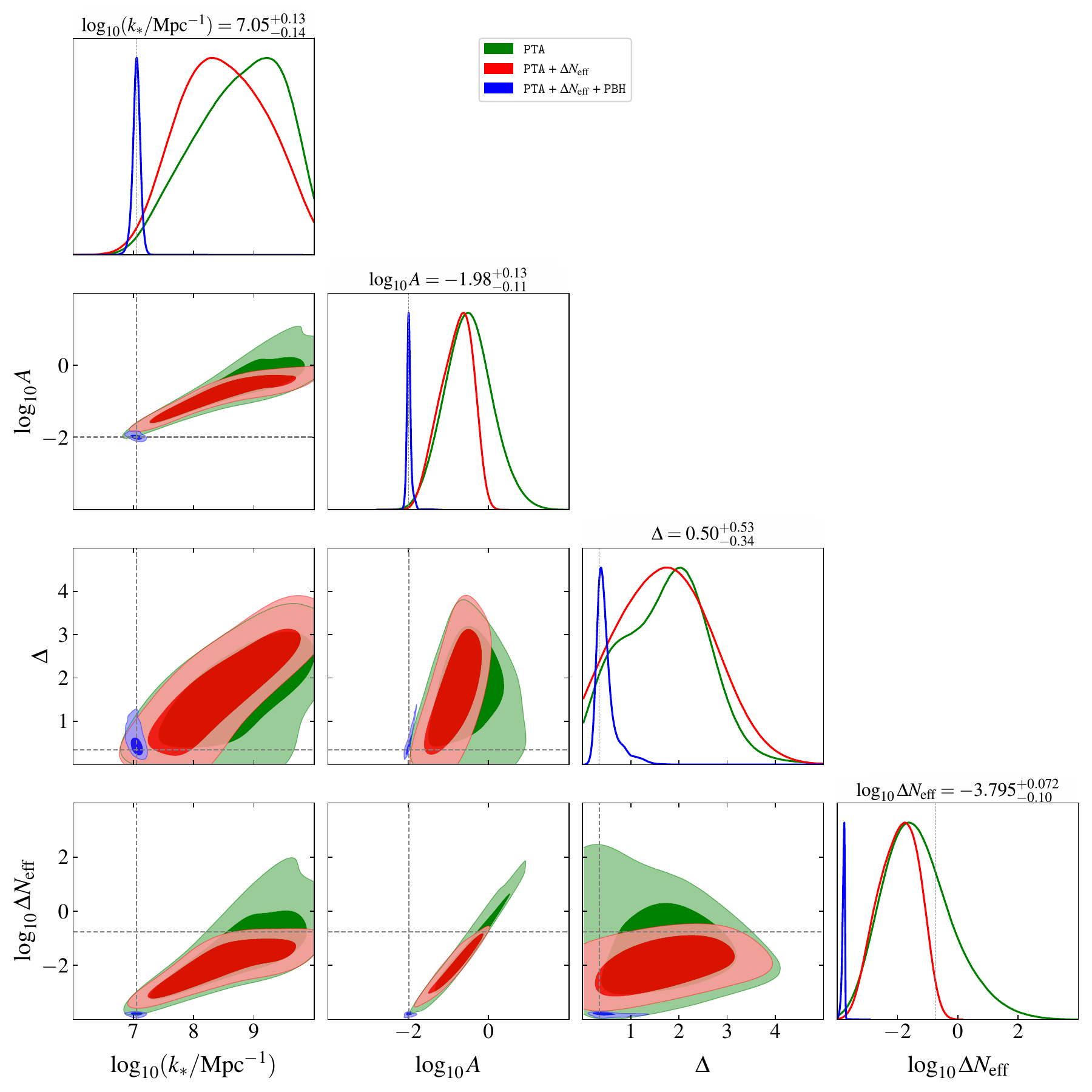}} \subfigure{\includegraphics[width=8.5cm]{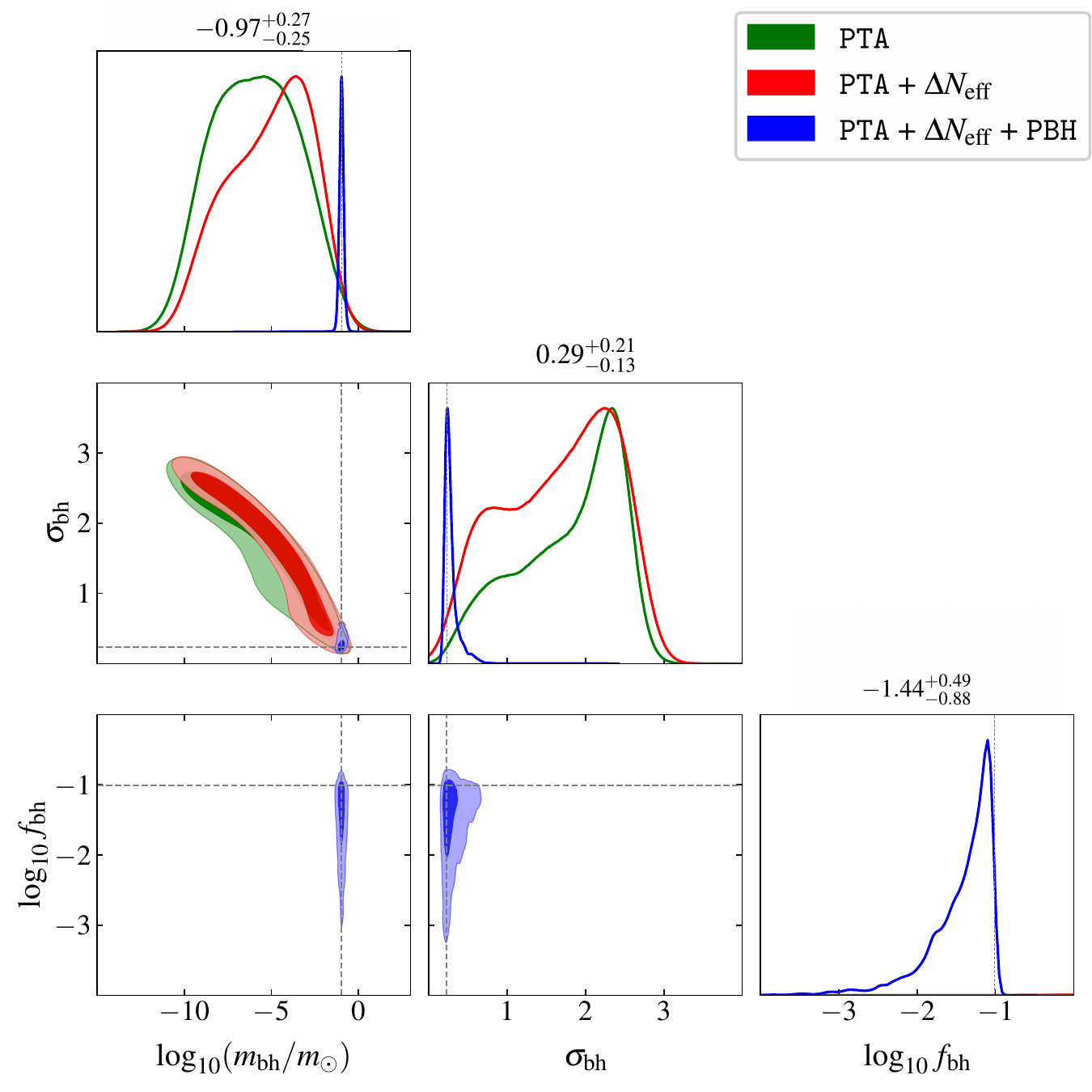}}\\
\caption{
Marginalised posteriors for $\ps$ parameters ($k_*,A,\Delta$) and derived $\dneff$ and PBH parameters ($m_\r{bh}, \fbh, \sbh$).
The green, red and blue contours correspond to {\tt PTA}, {\tt PTA+$\dneff$} and {\tt PTA+$\dneff$+PBH} inference settings respectively.
Light and dark-shaded regions correspond to 68\% and 95\% confidence levels respectively.
In $\dneff$ panels the dotted line indicates PLK upper limit of $\dneff < 0.175$,
and in other panels dotted lines mark the best-fit values,
numbers on diagonal panels show the marginalized mean and 95\% C.L. regions from our main {\tt PTA+$\dneff$+PBH} setting.
Note that $\fbh$ posteriors for {\tt PTA} and {\tt PTA+$\dneff$} settings are peaked at $\fbh > 10^{10}$ and are thus not shown in the figure.
}
\label{e2ftssaasadwu}
\end{figure*}

\section{Results}
\label{e2f8sanb_sfax34354tghr87t6eftcshidftyyrtdasadwu}

We sample $\ps$ parameter space with the likelihood in \Eq{dshf3765rghfdv} using {\tt multinest} sampler~\cite{Feroz:2008xx},
and we compute constraints for our $\ps$ parameters and various derived observables (e.g. $\dneff$, PBH parameters and merger GW) by analyzing  {\tt multinest}
chains using {\tt GetDist} package~\cite{Lewis:2019xzd}.
Table~\ref{tabd76543wedfghsa} summarizes prior ranges for our parameters along with their marginalized confidence region.
\Fig{e2ftssaasadwu} presents marginalized posteriors from our inference, 
in which the left panel shows results for our $\ps$ model parameters and the derived $\dneff$ from different inference settings, 
and the right panel shows results for PBH parameters $\fbh$, $m_\r{bh}$ and $\sigma_\r{bh}$.

As illustrated in the left panel of \Fig{e2ftssaasadwu} and from Table.\ref{tabd76543wedfghsa},
using PTA data alone gives a marginalised $\dneff$ 95\% C.L upper bound of 7.71,
which violates PLK upper limit by more than one order of magnitude.
{\tt PTA} posteriors also show positive correlation between $\ps$ amplitude $A$ and peak $k_*$,
and adding the prior in \Eq{dsfhi38r767fdg} helps breaking this degeneracy and tightens $\ps$ constraints.
For both {\tt PTA} and {\tt PTA + $\dneff$} settings,
the derived posterior for $\fbh$ can reach beyond the physically forbidden region of $\fbh\ge1$ by more than 13 orders of magnitudes,
and enforcing the physically motivated $\fbh < 0.1$ prior further reduces degeneracy between $A$ and $k_*$ and drastically tightens constraints on both $\ps$ and PBHs.
At 95\% C.L.,
PTA data alone constrains $m_\r{bh}$ and $\sbh$ to $[4.6 \times 10^{-11}, 3.9 \times 10^{-2}]\ m_\odot$ and $[0.48,\ 2.75]$ respectively.
In {\tt PTA+$\dneff$+PBH} fitting,
the PBH constraints significantly tightens to $[6.0 \times 10^{-2}, 2.0 \times 10^{-1}]\ m_\odot$ for $m_\r{bh}$ and $[0.16,\ 0.50]$ for $\sbh$,
whereas $\ps$ parameter posteriors shrink to $\log_{10}(k_*/\r{Mpc^{-1}}) = 7.05^{+0.13}_{0.14}$,
$\log_{10} A = -1.98^{+0.13}_{-0.11}$ and $\Delta = 0.50 ^{+0.53}_{-0.34}$.

\begin{figure*}[htp]
\centering
\subfigbottomskip=-500pt
\subfigure{
\subfigure{
\includegraphics[width=8.9cm]{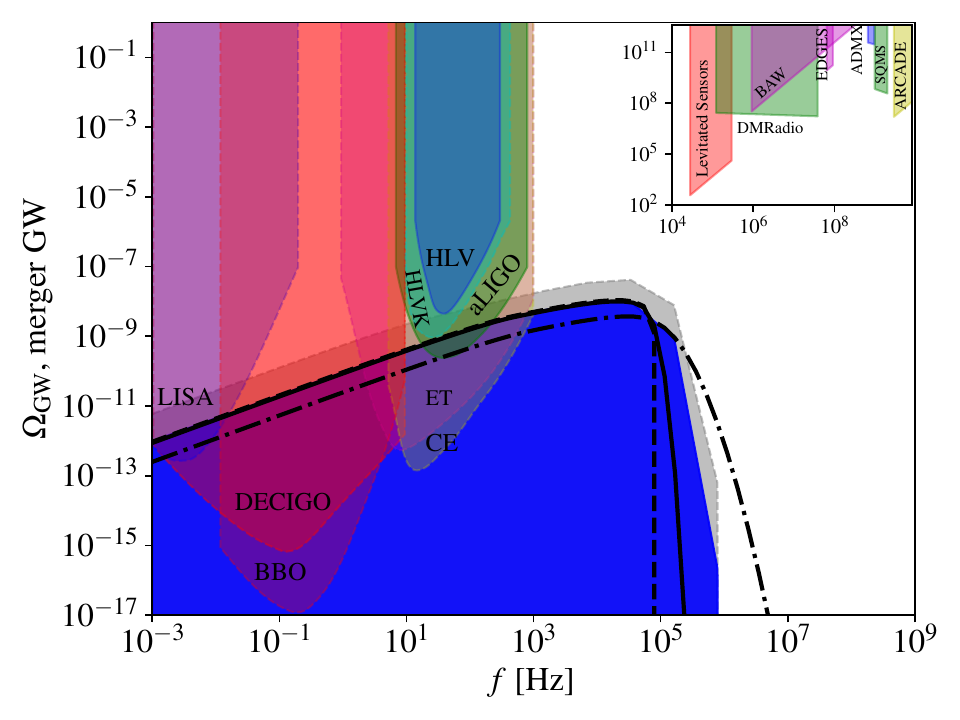}}
\includegraphics[width=9.1cm]{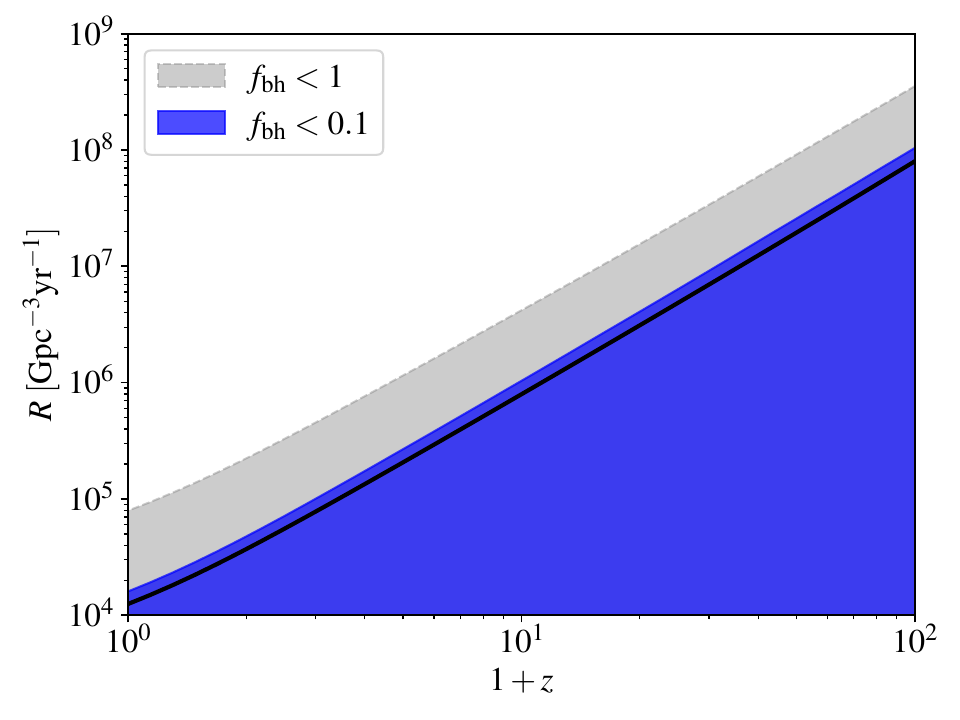}}
\caption{
95\% C.L. confidence region for merger $\Omega_\r{GW}$ (left) and merger rate (right),
legend applies to both panels.
The blue contours show results from {\tt PTA+$\dneff$+PBH} inference which enforces the $\fbh < 0.1$ prior,
we also show results when this prior is lifted to $\fbh < 1$,
however for $\fbh > 0.1$ the robustness of our merger calculation is yet to be verified~\cite{Raidal:2018bbj,Hall:2020daa},
thus we mark these regimes with grey contours.
The solid black lines correspond to our best-fit PBH parameters of $\fbh = 10^{-1},\ m_\r{bh} = 1.1 \times 10^{-1}m_{\odot},\ \sbh = 0.23$.
In the left panel,
we also show $\Omega_\r{GW}$ computed for $\sbh = 0$ (monochromatic PBHs, dashed) and $\sbh = 1$ (dot-dashed),
with the $\fbh$ and $m_\r{bh}$ values fixed to 0.1 and $1.1 \times 10^{-1}m_{\odot}$ respectively.
We also illustrate the landscape of experimental sensitivities from various GW observatories operating in our frequency window.
The experimental reach for 
LISA~\cite{LISA:2017pwj},
DECIGO~\cite{Kawamura:2011zz},
BBO~\cite{phinney2004big},
ET~\cite{Punturo:2010zz},
CE~\cite{Reitze:2019iox},
aLIGO~\cite{LIGOScientific:2014pky},
HLVK~\cite{NANOGrav:2023hvm,LIGOScientific:2014pky,VIRGO:2014yos,KAGRA:2018plz} and HLV~\cite{NANOGrav:2023hvm}
are the power law integrated sensitivities~\cite{Thrane:2013oya} from Refs.\cite{Thrane:2013oya, Domenech:2021ztg, Garcia-Bellido:2021jlq,NANOGrav:2023hvm}.
The sensitivities for
Levitated Sensors~\cite{Arvanitaki:2012cn},
bulk acoustic wave (BAW)~\cite{Goryachev:2014yra},
DMRadio~\cite{Silva-Feaver:2016qhh},
EDGES~\cite{Bowman:2018yin},
ADMX~\cite{ADMX:2021nhd},
SQMS~\cite{Herman:2020wao,Berlin:2021txa, Berlin:2023grv}
and ARCADE~\cite{Fixsen:2009xn} (shown in the inset) are adapted from~\cite{Franciolini:2022htd}.
We adopt design sensitivity for HLVK and third observation run for HLV~\cite{NANOGrav:2023hvm}.
Existing and projected experimental limits are indicated by solid and dashed edges, respectively.
}
\label{e2f8nb_asadwu}
\end{figure*}

SIGW interpretation of PTA signal can be further tested by GW missions operating at higher frequencies.
In blue-colored regions of \Fig{e2f8nb_asadwu},
we show 95\% C.L. posterior for merger GW (left) and merger rate (right) derived from our main {\tt PTA+$\dneff$+PBH} inference,
the black solid curves correspond to the maximum likelihood best-fit values of $\fbh = 0.1$, $m_\r{bh} = 1.1 \times 10^{-1} m_\odot$ and $\sbh = 0.23$.
Our $\Omega_\r{GW}$ posterior peaks at around 0.1 MHz with an amplitude of $\mathcal{O}(10^{-8})$,
towards lower frequencies $\Omega_\r{GW}$ decays as $f^{2/3}$,
and below $10^{4}$ Hz,
$\Omega_\r{GW}$ starts to fall into the sensitivity reach of various proposed experiments~\cite{Thrane:2013oya, Domenech:2021ztg, Garcia-Bellido:2021jlq},
such as aLIGO (Advanced LIGO)~\cite{LIGOScientific:2014pky},
LISA (Laser Interferometer Space Antenna)~\cite{LISA:2017pwj},
DECIGO~\cite{Kawamura:2020pcg},
ET~\cite{Punturo:2010zz} and BBO~\cite{phinney2004big}.
These missions are expected to achieve high-precision measurements through ground or space based laser Michelson interferometers,
the relevant arm lengths span from 4 km to $2.5\times10^6$ km~\cite{LIGOScientific:2014pky,LISACosmologyWorkingGroup:2022jok} and thereby covering a wide frequency range of $[10^{-4}, 10^3]$ Hz.
For DECIGO and BBO in particular,
$\Omega_\r{GW}$ posterior strength exceeds the sensitivity reach by about a factor of $10^4$ and $10^6$ respectively,
indicating a positive prospect for experimental searches.
The right panel of \Fig{e2f8nb_asadwu} shows that our inference constrains merger rate today to $R \lesssim 1.6 \times 10^4\ \r{Gpc^{-3}yr^{-1}}$.

In both panels, 
we also show results with a relaxed prior $\fbh < 1$ (grey),
equivalent to the requirement that PBH does not exceed the total dark matter density, 
with the caveat that our $\bar{N}(y)$ formulation needs validation for $f_{\rm bh} >0.1$~\cite{Raidal:2018bbj,Hall:2020daa}.
Fig.~\ref{e2f8nb_asadwu} shows that in this relaxed case, 
$\Omega_{\rm GW}$ posterior increases by about a factor of 4, 
and GW signals fall into the sensitivity reach of HLVK~\cite{NANOGrav:2023hvm,LIGOScientific:2014pky,VIRGO:2014yos,KAGRA:2018plz}
(a detector network consisting of aLIGO in Hanford and Livingston~\cite{LIGOScientific:2014pky},
aVIRGO~\cite{VIRGO:2014yos},
KAGRA~\cite{KAGRA:2018plz})
and HLV~\cite{NANOGrav:2023hvm} (similar to HLVK but without KAGRA),
whereas posterior for merger rate $R$ is raised by about a factor of 3.

While a high frequency signal would strongly indicate for a possible connection to the merger mechanism,  
it is interesting to note that even in case of null detection at high frequencies, there is still profound implication for PBHs, curvature perturbation and PTA signal interpretation.
We showcase this with another inference,
which will be dubbed as the {\tt Null Detection} inference hereafter.
Compared with the {\tt PTA+$\dneff$+PBH} setting,
we added an additional prior which sets the likelihood to zero when PBHs from sampled $\ps$ spectrum produce merger GW that falls into detectable regions of high frequency missions.

\begin{figure}[t]
\centering
\includegraphics[width=8.3cm]{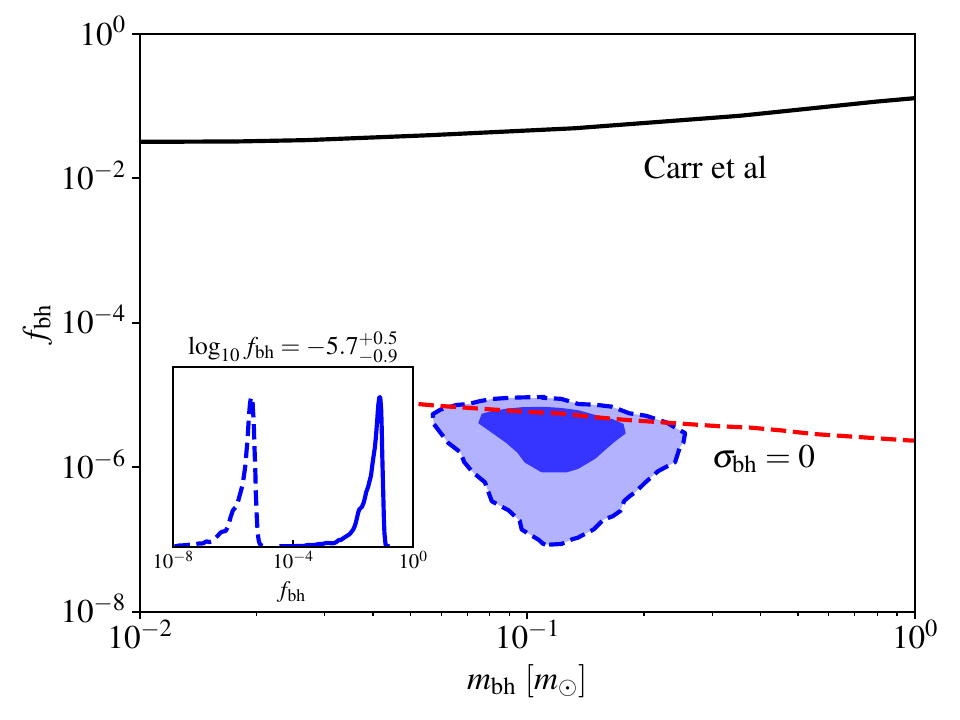}
\caption{
The implication of null detection of high frequency GWs on PBH abundance.
The blue shaded contours show marginalized 68\% C.L. (dark) and 95\% C.L. (light) regions,
obtained by adding the high frequency null-detection prior to the {\tt PTA+$\Delta N_\r{eff}$+PBH} inference.
The black solid curve shows the current upper limit on $\fbh$ summarized in Carr et al~\cite{Carr:2020xqk} for monochromatic PBHs.
The inset shows marginalized 1D distributions for $\fbh$,
with dashed and solid lines indicating {\tt Null Detection} and {\tt PTA+$\Delta N_\r{eff}$+PBH} inferences respectively.
The title of inset shows marginalized mean and 95\% C.L. region of $\fbh$ from {\tt Null Detection} inference.
Note that the blue shaded region does not correspond to any specific PBH distribution width $\sbh$ since it has been marginalized,
and in the red dashed line,
we showcase the expected {\tt Null Detection} upper limit for a monochromatic distribution ($\sbh = 0$).}
\label{e2f8sanb_asadwdssvdu}
\end{figure}

It has been previously shown in Ref.~\cite{Wang:2019kaf} that null detection of PBH merger GW can lead to very stringent constraints on PBH abundance,
and in \Fig{e2f8sanb_asadwdssvdu} we show joint posterior distribution for $\fbh$ and $m_\r{ bh}$ from {\tt Null Detection} inference.
For comparison,
we also show current $\fbh$ upper limits summarized in Ref.~\cite{Carr:2020xqk} for monochromatically distributed PBHs.
Such distribution assumes that all PBHs have the same mass and is a limiting case of $\sbh \to 0$.
Currently around $[6 \times 10^{-2}, 0.2]m_\odot$ mass window favored by PTA data,
$\fbh$ is primarily constrained by gravitational lensing observations~\cite{Carr:2020xqk,EROS-2:2006ryy,Niikura:2017zjd} to $\fbh \lesssim 10^{-1}$,
and it can be seen from the upper boundary of our posterior contour that,
for majority of this mass range,
PBHs can be detected or excluded (in event of null detection) if their abundance exceeds $\fbh \sim 10^{-5}$.
Such constraints are stronger than existing ones by about 4 orders of magnitudes.
PBHs are seriously over-produced with SIGW interpretation of PTA data,
and 1D marginalized $\fbh$ posterior in the inset of \Fig{e2f8sanb_asadwdssvdu} demonstrates that,
while posterior from {\tt PTA+$\dneff$+PBH} setting (blue solid) is peaked at the upper edge of our $\fbh$ prior,
{\tt Null Detection} drastically lowers $\fbh$ posterior by 5 orders of magnitudes.

Note our confidence contour in \Fig{e2f8sanb_asadwdssvdu} does not correspond to any specific value of $\sbh$ because $\sbh$ has been marginalized over.
For a more rigorous comparison with existing monochromatic PBH bounds,
in the red dashed line we present our sensitivity estimate for monochromatic PBHs ($\sbh=0$).
We obtained this constraint by iteratively solving for the maximum $\fbh$ value whose merger GW escapes the high frequency sensitivity reach,
and it can be seen that our $\fbh$ upper limit scales as $m_\r{bh}^{-0.4}$ and reaches $\fbh \lesssim 6\times 10^{-6}$ at $0.1 m_\odot$,
improved by about 4 order of magnitudes compared with existing limits.

\begin{figure}[htp]
\centering
\includegraphics[width=8.5cm]{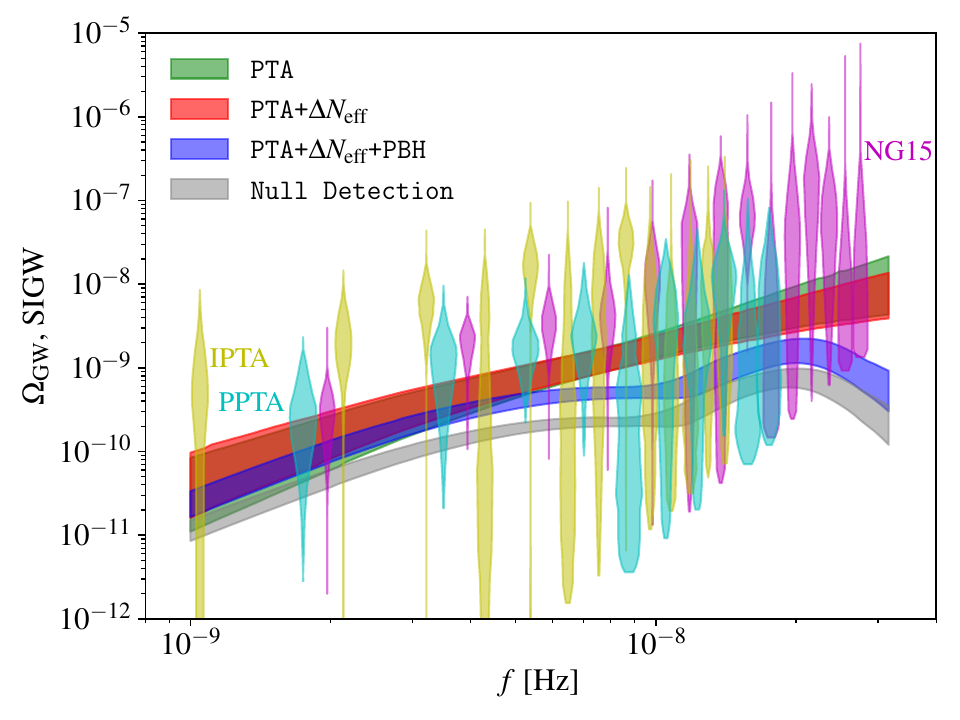}
\caption{
Posterior distributions of SIGW and comparison with experimental data.
The magenta, yellow and cyan regions show the measured $\GW$ posterior from NANOGrav 15 years result (NG15)~\cite{NANOGrav:2023gor},
IPTA~\cite{Antoniadis:2022pcn} and PPTA~\cite{Reardon:2023gzh} respectively.
The green, red, and blue contours show our 68\% C.L. posterior for SIGW from {\tt PTA}, {\tt PTA+$\dneff$} and {\tt PTA+$\dneff$+PBH} settings,
and the grey contour shows 68\% C.L. posterior from {\tt Null Detection} inference,
in which we apply to {\tt PTA+$\dneff$+PBH} setting an additional prior that sets likelihood to zero when PBHs produced from sampled $\ps$ generate GW reachable by high frequency experiments.
}
\label{e2f8sanb_asadwu}
\end{figure}

\Fig{e2f8sanb_asadwu} summarises the comparison between PTA datasets and $\GW$ posteriors from different inferences.
$\GW$ from {\tt PTA} and {\tt PTA+$\dneff$} settings both show good agreement with the data.
For {\tt PTA+$\dneff$+PBH} inference,
the $\fbh$ prior requires a significantly lower $\ps$ amplitude,
therefore the relevant $\GW$ posterior is shifted below that of {\tt PTA} and {\tt PTA+$\dneff$} runs.
In order not to overproduce PBHs detectable through high frequency merger GW,
{\tt Null Detection} setting requires even lower $\fbh$,
which further suppresses the amplitude of $\ps$ responsible for PBH production and thereby driving SIGW posterior even lower.
These visible shifts in SIGW posterior can be useful in discriminating PTA candidate sources,
which is currently a highly debated issue~\cite{NANOGrav:2023hvm}.

As expected,
from \Fig{e2f8sanb_asadwu} one can see that the {\tt Null Detection} inference,
which enforces non-detection of high frequency GW signal predicted by joint {\tt PTA+$\dneff$+PBH} likelihood,
gives a worse fit to PTA data compared to other inference settings.
In {\tt PTA+$\dneff$+PBH} fitting we obtain a minimum $\chi^2$ of $\chi^2_\r{min} = 19.1$
\footnote{
We define $\chi^2_\r{min}$ for an inference as,
\be
\chi^2_\r{min}
\equiv
\min
\left(
\sum_{i}
\frac{(x_i - u_i)^2}{\sigma_i^2}
\right)
\ee
where the minimization is performed over all parameter samples,
$x_i$, $u_i$ and $\sigma_i$ refer to model prediction,
data median and error for $\ln \GW$ as used in \Eq{dshf3765rghfdv}.
},
whereas in {\tt Null Detection} case $\chi^2_\r{min}$ increases dramatically to 36.9,
indicating that non-detection of high frequency PBH merger GW may imply serious tension with PTA observations if the PTA signals are indeed sourced by SIGW,
and this can further help determining PTA sources.

Finally we specify the main differences between our analysis and several studies that appeared earlier in the literature.
In Ref.~\cite{Kohri:2020qqd},
the authors pointed out that merger GW from PBHs specifications indicated by NANOGrav 12.5-year dataset is potentially detectable by high frequency observations,
and Ref.~\cite{Inomata:2023zup} updated the analysis for the latest NANOGrav 15-year observation with several representative PBH specifications.
Our paper can be considered as a rigorous and comprehensive extension. 
Using the 1$\sigma$ and 2$\sigma$ posterior contours for GW density or $\ps$ parameters given by inferences performed in NANOGrav collaboration papers~\cite{NANOGrav:2020bcs,NANOGrav:2023hvm},
Refs.~\cite{Kohri:2020qqd,Inomata:2023zup} picked and computed the relevant merger GW,
thereby showcasing the feasibility of cross-linking PTA GW with high frequency observations.
Here we adopted a more rigorous analysis with our inferences,
and we used the inference chains to map the posteriors for $\ps$ parameters into posteriors for PBHs and merger GW,
therefore in addition to signals for representative scenarios,
\Fig{e2f8nb_asadwu} also includes the full Bayesian posterior for merger GW which covers the entire posterior region of parameter space. 
We work with realistic scenarios covering the entire probable parameter space,
thereby avoiding possible bias for PBH and merger GW posteriors.
As discussed in our previous texts and also in~\cite{Kohri:2020qqd,Zhao:2022kvz,NANOGrav:2023hvm,Inomata:2023zup},
the PTA data alone is known to seriously over-produce PBHs,
and here in addition to including more updated PTA datasets (IPTA and PPTA), 
our likelihood priors also take into account the the existing constraints on PBH and $\dneff$.
These additional priors have a significant impact on inference results and helps excluding unphysical models that over-produce PBHs or $\dneff$.
Our investigation of the implication of null detection of high frequency GW represents another modeling improvement of this work.

\section{Discussions}
\label{e2f8sanb_sfax34354tghr87t6eftcs_dsfeu}

Enhanced curvature perturbation $\ps$ can produce SIGW which serves as a good source candidate of the GW signal recently reported by various PTA experiments.
This paper scrutinizes the implication of this scenario at the higher frequency band of GW spectrum.
We perform extensive inference analysis of PTA GW datasets in combination with existing constraints on integrated GW density (parameterized by $\dneff$) and PBH abundance,
from which we map out posterior distributions for PBHs and relevant merger GW signals.
Our result shows that if the PTA GW signals are indeed sourced by SIGW,
the $\ps$ amplitude required will create PBHs in a narrow $[6 \times 10^{-2}, 2 \times 10^{-1}]\ m_\odot$ mass range.
Mergers of these PBHs will produce strong GW background across $[10^{-3}, 10^5]$ Hz frequencies,
which falls into sensitivity reach of various GW projects such as LISA,
aLIGO,
DECIGO,
CE,
ET and BBO,
thus high frequency GW experiments can help further testing the SIGW scenario for PTA gravitational waves and potentially improve current PBH constraints by up to 5 orders of magnitudes.
We also show that PTA data significantly overproduce PBHs,
and incorporating current PBH abundance constraints leads to visible shifts in SIGW posteriors which can help discriminating PTA candidate sources.

\bigskip
{\bf Acknowledgements.}
The authors thank Gabriele Franciolini and Paul Frederik Depta for helpful discussions. This work is supported by the National Natural Science Foundation of China (No. 12105013 and No. 12275278). 

\appendix
\section{Comparison of approximate GW likelihood with full analysis}
\label{appendix_Comparison_with_Full_GW_Analysis}

\begin{figure}[b]
\centering
\includegraphics[width=8.3cm]{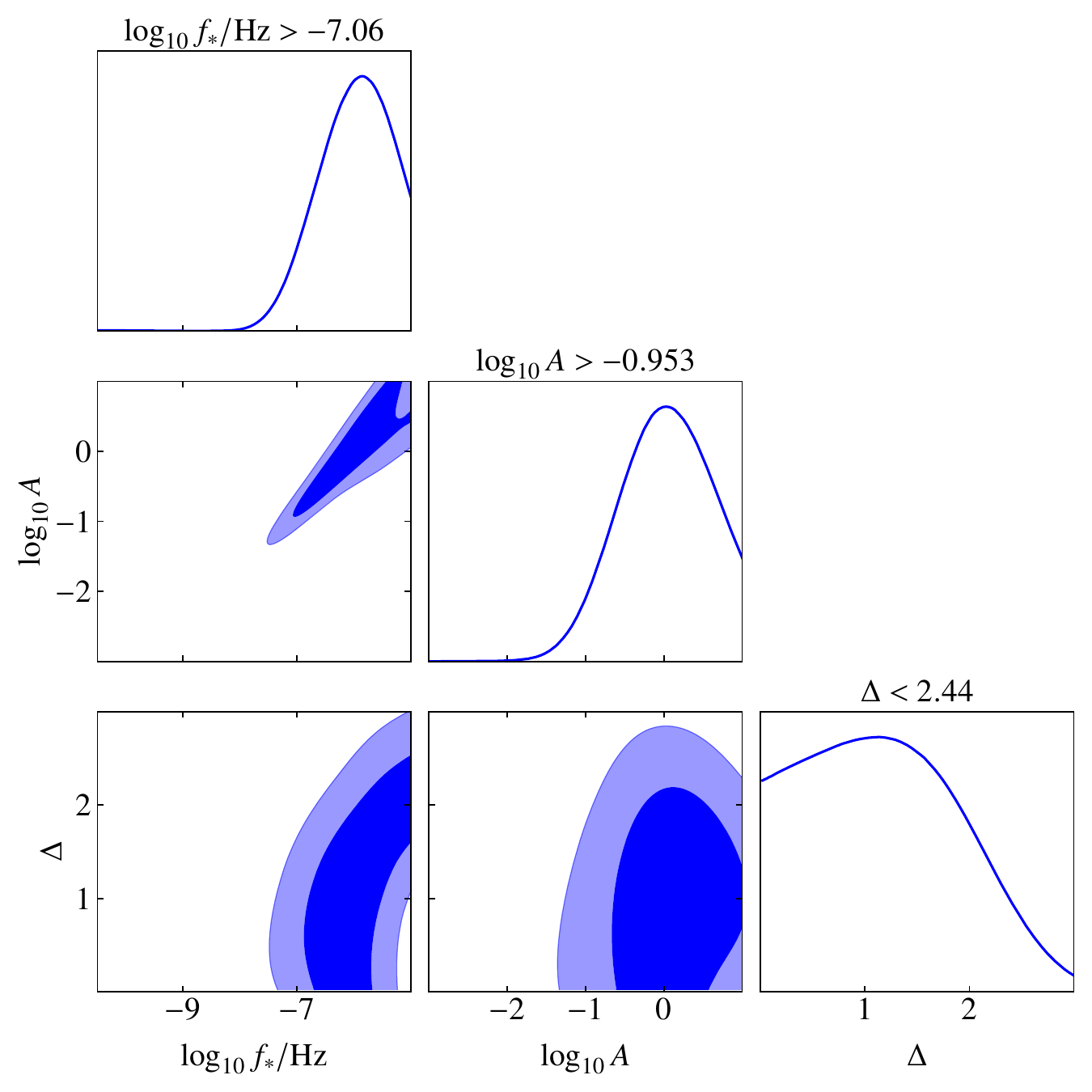}
\caption{
Posterior distributions for $\ps$ parameters $f_*$, $A$ and $\Delta$,
derived by sampling NAGOGrav-15 GW data~\cite{NANOGrav:2023hvm} using the approximate likelihood in \Eq{dshf3765rghfdv}.
The dark and light shaded regions show the 68\% and 95\% C.L.,
titles above each panel show the 95\% C.L..
As a comparison,
our result here using approximate GW likelihood shows good agreements with the full NAGOGrav-15 analysis shown in Fig.7b of Ref.~\cite{NANOGrav:2023hvm}.
}
\label{Fig_NG15_CornorPlot}
\end{figure}

Our approximate GW likelihood detailed in \Eq{dshf3765rghfdv} had been cross-checked with Refs.~\cite{NANOGrav:2023hvm,Franciolini:2023pbf} to give consistent results.
As a comparison with the full GW analysis in \cite{NANOGrav:2023hvm} which used NAGOGrav-15 data,
here in \Fig{Fig_NG15_CornorPlot} we show the posterior distributions for $\ps$ parameters $f_*$, $A$ and $\Delta$ derived from our approximate GW likelihood,
where $f_*$ is related to $k_*$ in \Eq{dsf9887hdsf} through \Eq{eq_k2f},
\be
f = 1.546 \times 10^{-15}\left(\frac{k_*}{\r{Mpc^{-1}}}\right) \r{Hz}.
\ee
For consistency,
we use only the NAGOGrav-15 data in our analysis as in \cite{NANOGrav:2023hvm},
and we adopt the same prior ranges as in~\cite{NANOGrav:2023hvm}.

Comparing with Fig.7b of \cite{NANOGrav:2023hvm},
\Fig{Fig_NG15_CornorPlot} shows that when using the same GW data and parameter prior,
our approximate GW likelihood shows very close match with the full analysis in \cite{NANOGrav:2023hvm}.

\bibliography{main.bib}

\end{document}